\documentclass[prd,onecolumn,showpacs,superscriptaddress,nofootinbib,floatfix,showkeys,10pt]{revtex4-2}
\usepackage{graphicx}
\usepackage{amsmath}
\usepackage{bm}
\usepackage{yhmath}
\usepackage{mathtools}
\usepackage{wasysym}
\usepackage[colorlinks,citecolor=blue,urlcolor=blue,linkcolor=blue]{hyperref}
\usepackage{subfigure}
\usepackage{color}
\usepackage{cases}
\usepackage{subfigure}
\usepackage{times}
\usepackage{dcolumn,booktabs,bm}
\usepackage{slashed}
\usepackage{amsfonts,amssymb,stmaryrd,latexsym,amsmath}
\usepackage{textcomp}
\usepackage{multirow}
\usepackage{cancel}
\usepackage{array}
\usepackage{orcidlink}
\usepackage{enumitem}
\usepackage{float}

\renewcommand{\arraystretch}{1.8}

\begin{document}

\title{Masses and Magnetic Moments of Doubly Heavy Tetraquarks via Diffusion Monte Carlo Method}

\author{Halil Mutuk}
 \email{hmutuk@omu.edu.tr}
 \affiliation{Department of Physics, Faculty of Science, Ondokuz Mayis University, 55200 Samsun, Türkiye}

\begin{abstract}
We present mass spectrum and magnetic moments of the $\bar{n}\bar{n}QQ$ states, where $n=u,d,s$ and $Q=c,b$. We solve four-body Schrödinger equation with a quark potential model by using diffusion Monte Carlo (DMC) method.
The quark potential is based on the Coulomb, confinement and spin-spin interaction terms. We find mass and magnetic moment of the $T_{cc}^+$ state as $M_{T_{cc}^+}=3892 ~\text{MeV}$ and $\mu=0.28 \mu_N$,respectively. We also find the mass and magnetic moment of $T_{bb}^-$ as $M_{T_{bb}^-}=10338 ~\text{MeV}$ and $\mu=-0.32 \mu_N$, respectively. We find some bound state candidates of doubly heavy tetraquark systems with $I(J^P)=0(1)^+$ $nn \bar b \bar b$, $I(J^P)=0(0)^+$ $nn \bar c \bar b$, $I(J^P)=0(1)^+$ $nn \bar c \bar b$, and $I(J^P)=1/2(1)^+$ $ns \bar b \bar b$. We compare our results with other approaches in the literature.
\end{abstract}

\maketitle

\section{Prologue}
After the breakthrough works of Gell-Mann \cite{Gell-Mann:1964ewy} and Zweig \cite{Zweig:1964jf} in 1964, hadrons can be classified into two groups in terms of their valence quarks and antiquarks: quark-antiquark $(q \bar q)$ configuration as mesons and three-quark configuration $(qqq)$ as baryons. This classification was called as \textit{quark model} and reproduced reasonably well many hadron observables such as baryon and meson spectrum, strong decays, magnetic moments of baryons, etc. However, the Quantum Chromodynamics (QCD) allows exotic hadrons beyond the scope of quark model. The existence of multiquark states is consistent with QCD. In fact, in Refs. \cite{Gell-Mann:1964ewy,Zweig:1964jf}, multiquark states beyond conventional ones (mesons and baryons) were conjectured as combinations of quarks. The era of conventional hadrons directed in a different route when the observation of $\chi_{c1}(3872)$ by the Belle Collaboration \cite{Belle:2003nnu} announced in 2003. This state has a narrow width, decays into $J/\psi 2 \pi$ and $J/\psi 3 \pi$ and does not fall into the traditional charmonium spectra. Many unconventional  states have been reported by Belle, BaBar, CLEO, BESIII, CDF, D0, LHCb, ATLAS, and CMS collaborations since 2003. These unconventional charmonium and bottomonium states are named as XYZ mesons. For a recent naming convention of these states, see Ref. \cite{Gershon:2022xnn}. These unconventional states mostly decay into final states which include $(c \bar c)$ and $(b \bar b)$ pairs.  At this point, heavy quarkonium physics becomes important for elucidating these states. The current status and prospects of the heavy quarkonium physics can be found in Refs. \cite{Lebed:2016hpi,Ali:2017jda,Olsen:2017bmm,Guo:2017jvc,Liu:2019zoy,Brambilla:2019esw}.

The recent discovery of $T_{cc}^+$ tetraquark is of particular importance due to the its quark content \cite{LHCb:2021vvq,LHCb:2021auc}. The quark configuration is $cc\bar{u}\bar{d}$ and its mass is a few MeV below the lowest two-meson threshold of $D D^\ast$. Since the lowest comparable decay threshold consists of heavy-light meson pairings rather than a threshold consisting of a light meson and quarkonium, such $QQ \bar q \bar q$ multiquark states are flavor unusual. Exotic hadrons with two heavy quarks, $QQ \bar q \bar q$, were in the to do's list starting from 1980s \cite{Ader:1981db, Ballot:1983iv, Lipkin:1986dw, Zouzou:1986qh, Heller:1986bt, Carlson:1987hh}. Especially during the last decade, doubly heavy tetraquarks were studied by means of quark level models \cite{Silvestre-Brac:1993zem, Semay:1994ht, Chow:1994mu, Chow:1994hg, Pepin:1996id,Brink:1998as, Gelman:2002wf, Vijande:2003ki, Janc:2004qn, Vijande:2006jf, Ebert:2007rn, Zhang:2007mu, Vijande:2007ix, Vijande:2009kj, Yang:2009zzp, Feng:2013kea, Karliner:2017qjm, Eichten:2017ffp, Yan:2018gik, Ali:2018xfq, Park:2018wjk, Deng:2018kly, Carames:2018tpe, Hernandez:2019eox, Yang:2019itm, Bedolla:2019zwg, Yu:2019sxx, Wallbott:2020jzh, Tan:2020ldi, Lu:2020rog, Braaten:2020nwp, Yang:2020fou, Giron:2020qpb, Qin:2020zlg, Meng:2020knc, Meng:2021yjr, Chen:2021tnn, Jin:2021cxj, Giron:2021sla, Chen:2021cfl, Andreev:2021eyj, Deng:2021gnb, Ortega:2022efc, Kim:2022pyq}, chromomagnetic interaction models \cite{Lee:2007tn, Lee:2009rt, Hyodo:2012pm, Luo:2017eub, Hyodo:2017hue, Cheng:2020nho, Cheng:2020wxa, Weng:2021hje, Guo:2021yws}, QCD sum rule \cite{Navarra:2007yw, Dias:2011mi, Du:2012wp, Chen:2013aba, Wang:2017uld, Wang:2017dtg, Agaev:2018khe, Sundu:2019feu, Agaev:2019kkz, Tang:2019nwv, Agaev:2019lwh, Agaev:2020dba, Wang:2020jgb, Agaev:2020zag, Agaev:2020mqq, Agaev:2021vur, Azizi:2021aib, Aliev:2021dgx, Ozdem:2021hmk, Bilmis:2021rdp,Ozdem:2022yhi,Azizi:2023gzv}, lattice QCD simulations \cite{Detmold:2007wk, Wagner:2010ad, Bali:2010xa, Bicudo:2012qt, Brown:2012tm, Ikeda:2013vwa, Bicudo:2015vta, Bicudo:2015kna,  Francis:2016hui, Bicudo:2016ooe, Bicudo:2017szl, Francis:2018jyb, Junnarkar:2018twb, Leskovec:2019ioa, Hudspith:2020tdf, Mohanta:2020eed, Bicudo:2021qxj, Padmanath:2021qje,Hoffmann:2022jdx,Meinel:2022lzo}, and so on. $QQ \bar q \bar q$ tetraquarks should be stable when the quark masses are sufficiently large. In other words, when the ratio of $m_{QQ}/m_{qq}$ is large in the four-quark state, then it is natural to consider the stability of this multiquark state \cite{Carlson:1987hh,Manohar:1992nd}. In this limit two heavy quarks (antiquarks) form an anticolor (color) triplet due to the Coulomb term of the potential at small $QQ$ $(\bar Q \bar Q)$ separations. Then the four-quark structure is quite similar to $\bar Q \bar q \bar q$ $(Qqq)$ baryons. The same logic applies to the $(\bar Q \bar Q q)$ baryons which are related to heavy-light mesons $Q \bar q$ \cite{Savage:1990di,Brambilla:2005yk,Cohen:2006jg,Mehen:2017nrh}. Furthermore, heavy quark symmetry (HQS) also implies stable $QQ \bar q \bar q$ tetraquark states \cite{Eichten:2017ffp}.  

Hadron spectroscopy constitutes basis to test predictions of QCD. Hadron spectroscopy is related to masses and decays of the hadrons. The abundance and clear signals in the charmonium and bottomonium states prompt hadron spectroscopy. Charmonium and bottomonium states are heavy quarkonium states. Heavy quarkonium is a good laboratory for QCD: at high energies (small distances) where a perturbative calculations are possible and at low energies (large distances) nonperturbative effects take place  \cite{Brambilla:2010cs}. Since charm and bottom quarks are heavy, nonrelativistic approach of QCD can be used for these mesons.

There are various methods to study physical properties of exotic states. QCD motivated phenomenological models can be used to classify and specify the components (coloured constituents) and forces between them in order to study these exotic states. In that line, many important problems related to these states can be addressed by solving Schrödinger equation. Since these exotic states include at least more than three quarks, solving many-body Schrödinger equation is quite challenging. In general four-quark states are often taken as diquark-antidiquark states or meson-meson molecules. Such quark clusterings yield a two-body Schrödinger equation in the context of potential models. In this present work we rely on a different method for solving four-quark Schrödinger equation. In our model, none of the valence quarks will be assumed to be clustered together. The interaction in the context of potential model for this system is the widely accepted and used one: Coulomb + linear confining + spin-spin term.  We solve four-quark Schrödinger equation by using Diffusion Monte Carlo (DMC) method which takes into account full correlations between the constituents of the system and is capable of producing exactly the lowest eigenenergy since the effect of trial wave function is diminished by the method.

The order of the paper is as follows: We describe the potential model, numerical approach in Sec. \ref{sec:level2}. We present and explain the mass spectrum and magnetic moments of the doubly heavy tetraquark states in Sec. \ref{sec:level3}. The conclusion is reserved for Sec. \ref{sec:level4}.

\section{\label{sec:level2}Theoretical Framework}
\subsection{Potential Model}
Phenomenological models with a simple relativistic kinetic energy term and a scalar potential term which incorporates the so called linear confinement, and a term related to short distance which incorporates color-Coulomb interaction stemmed from QCD, give good results and descriptions of the observed spectra of both heavy and light quark mesons and baryons. 

The Hamiltonian which describes doubly heavy tetraquark states  reads as follows \cite{SilvestreBrac:1996bg}
\begin{equation}
H=\sum_i \left( m_i+\frac{\bf{p}_i^2}{2m_i} \right)-\frac{3}{16}\sum_{i<j} \tilde{\lambda}_i \tilde{\lambda}_j v_{ij}(r_{ij})
\end{equation}
with the potential 
\begin{eqnarray}
v_{ij}(r)&=&-\frac{\kappa(1-e^{-\frac{r}{r_c}})}{r}+\lambda r^p + \Lambda \\ &+&
           \frac{2\pi}{3m_im_j}\kappa^\prime (1-e^{-\frac{r}{r_c}})\frac{e^{-\frac{r^2}{r_0^2}}}{\pi^{3/2}r_0^3} \bf{\sigma_i} \bf{\sigma_j}, \label{potential1}
\end{eqnarray}
where $r_0(m_i,m_j)=A \left(\frac{2m_im_j}{m_i+m_j} \right)^{-B}$, $A$ and $B$ are constant parameters, $\kappa$ and $\kappa^\prime$ are parameters, $r_{ij}$ is the interquark distance $\vert \bf{r_i}-\bf{r_j \vert}$, $\sigma_i$ are the Pauli matrices and $\tilde{\lambda}_i$ are Gell-Mann matrices. $\kappa$ and $\kappa^\prime$ are related to the hyperfine interactions, just like the replacement of $\delta(\vec{r})$ with a smearing function in the Breit-Fermi approximation. There are four potentials referred to the $p$ and $r_c$:
\begin{eqnarray*}
\text{AL1} &\to & p=1, ~ r_c=0, \\
\text{AP1} &\to & p=2/3, ~ r_c=0, \\
\text{AL2} &\to & p=1, ~ r_c \neq 0, \\
\text{AP2} &\to & p=2/3,  r_c \neq 0.
\end{eqnarray*}

This potential was developed under the nonrelativistic quark model (NRQM) and used for exploratory studies. It compose of ``Coulomb + linear" or ``Coulomb + 2/3-power'"  term and a strong but smooth hyperfine term. As mentioned in Ref. \cite{SilvestreBrac:1996bg}, potentials with $r_c \neq 0$ give results slightly better than  those with $r_c=0$, where 
$r_c=0$ means $\kappa(r)$ and $\kappa(r)^\prime$ are constant parameters. Therefore we choose AL2 potential with the following parameter listed in Table \ref{tab:table1}:

\begin{table}[H]
\caption{\label{tab:table1}Parameters of the AL2 potential}
\begin{ruledtabular}
\begin{tabular}{cc}
 &AL2\\
\hline
$m_u=m_d$&  0.320 GeV  \\
 $m_s$&  0.587 GeV  \\
$m_c$ &  1.851 GeV  \\
$m_b$ & 5.231 GeV \\
$\kappa$ & 0.5871  \\
$\kappa^\prime$ &  1.8475 \\
$\lambda$&  0.1673 $\text{GeV}^2$  \\
$\Lambda$&  -0.8182 GeV  \\
$B$ & 0.2132\\
$A$&1.6560 $\text{GeV}^{B-1}$ \\
$r_c$ & 0.1844 $\text{GeV}^{-1}$ 
\end{tabular}
\end{ruledtabular}
\end{table}

\subsection{Diffusion Monte Carlo Method}
Many important aspects of microscopic phenomena can be provided by the Schrödinger equation. Unfortunately, there are very few instances in which the Schrödinger equation can be solved exactly. For the rest which include more realistic cases, one needs to use approximations and numerical descriptions. Diffusion Monte Carlo (DMC) technique may help on this endeavour. 

The  imaginary-time Schrödinger equation (in $\hbar=c=1$) can be written as
\begin{equation}
-\frac{\partial \Psi_{\alpha'}(\bm{R},t)}{\partial t} = (H_{\alpha'\alpha}-E_R) \Psi_{\alpha}(\bm{R},t) \,,
\label{eq:Sch1}
\end{equation}
with the Hamiltonian defined as
\begin{equation}
H_{\alpha'\alpha} = -\frac{\nabla_{\bm{R}}^2}{2m} \, \delta_{\alpha'\alpha} + V_{\alpha'\alpha}(\bm{R}).
\end{equation}
In the above equations, $\bm{R}\equiv(\vec{r}_1,\ldots,\vec{r}_n)$  represents the positions of $n$ particles, $E_R$ is the shifted energy, and $\alpha$ denotes each possible spin-color channel of the system. The wave function in Eq. (\ref{eq:Sch1}) can be expanded as
\begin{equation}
\Psi_{\alpha}(\bm{R},t)=\sum_{i}c_{i,\alpha} \Phi_{i,\alpha}(\bm{R})e^{-[E_{i}-E_{R}]t},
\end{equation}
where $E_i$ are the eigenvalues and $\Phi_{i,\alpha}(\bm{R})$ are the eigenfunctions. If $E_R$ is taken close to the ground state energy $E_0$, i.e. $E_R \simeq E_0$, the wave function $\Psi_{\alpha}(\bm{R},t)$ will approach to $\Phi_{0,\alpha}(\bm{R})$ when  sufficiently long time elapses, i.e. $t\to \infty$. 

An important feature of the quantum monte carlo (QMC) method is to use sampling technique to reduce statistical fluctuation \cite{Kalos1974}. In this approach, we use the function
\begin{equation}
f_{\alpha}(\bm{R},t) \equiv \psi(\bm{R}) \, \Psi_{\alpha}(\bm{R},t) \,, \label{trifunc}
\end{equation}
where $\psi(\bm{R})$ is the time-independent trial function which in principle should be chosen as close to $\Phi_{0,\alpha}(\bm{R})$ as possible. Since $\psi(\bm{R})$ is a trial function, it should be chosen according to the sake of the calculations. In this present work, we formulate it as \cite{Gordillo:2020sgc}
\begin{equation}
\psi(\bm{R}) = \prod_{i<j} \Phi(\vec{r}_{ij})=\prod_{i<j} e^{-a_{ij}r_{ij}} \,,
\end{equation}
where the values of the adjustable constants $a_{ij}$ are set to minimize fluctuation. 

With this new definition of the trial function (Eq. \ref{trifunc}), the Schrödinger equation turns out to be 
\begin{eqnarray}
-\frac{\partial f_{\alpha'}(\bm{R},t)}{\partial t} &=& - \frac{1}{2m} \nabla_{\bm{R}}^2 f_{\alpha'}(\bm{R},t) + \frac{1}{2m} \nabla_{\bm{R}} \big[ F(\bm{R})f_{\alpha'}(\bm{R},t) \big] \nonumber \\ &+& \big[ E_{L}(\bm{R})-E_s \big] f_{\alpha'}(\bm{R},t) \nonumber \\ &+& V_{\alpha'\alpha}(\bm{R}) f_{\alpha}(\bm{R},t),\label{eq:Sch2}
\end{eqnarray}
where $E_{L}(\bm{R}) = \psi(\bm{R})^{-1} \, H_{0}\, \psi(\bm{R})$ is the local energy and  $F(\bm{R}) = 2\, \psi(\bm{R})^{-1}\, \nabla_{\bm{R}} \,\psi(\bm{R})$ is the drift force. The solution to Eq. (\ref{eq:Sch2}) in path integral formalism reads as
\begin{align}
f_{\alpha'}(\boldsymbol{R}',t+\Delta t) &= \int \mathrm{d}\boldsymbol{R_1}\mathrm{d}\boldsymbol{R_2}\mathrm{d}\boldsymbol{R_3}\mathrm{d}\boldsymbol{R_4}\mathrm{d}\boldsymbol{R} \nonumber \\
&
\hspace*{-2.20cm} \times \Bigg[ G^{(3)}\Big( \bm{R}',\bm{R}_1,\frac{\Delta t}{2}\Big) \, G^{(2)}\Big(\bm{R}_1,\bm{R}_2,\frac{\Delta t}{2}\Big) \nonumber \\
&
\hspace*{-2.20cm} \times G^{(1)}\Big( \bm{R}_2,\bm{R}_3,\Delta t \Big) \nonumber \\
&
\hspace*{-2.20cm} \times G^{(2)}\Big( \bm{R}_3,\bm{R}_4,\frac{\Delta t}{2}\Big) \, G^{(3)}\Big(\bm{R}_4,\bm{R},\frac{\Delta t}{2}\Big) \Bigg] \nonumber \\
&
\hspace*{-2.20cm} \times \sum_{\alpha}\, e^{-V_{\alpha'\alpha}(\bm{R}) \Delta t} f_{\alpha}(\bm{R},t) \,,
\end{align}
with
\begin{align}
&
G^{(1)}(\boldsymbol{R}',\boldsymbol{R},t) = \left(\frac{2\pi t}{m}\right)^{-\frac{3n}{2}} \, e^{\left(-\frac{m(\boldsymbol{R}'-\boldsymbol{R})^2}{2t}\right)} \,, \\[1ex]
&
G^{(2)}(\boldsymbol{R}',\boldsymbol{R},t) = \delta(\boldsymbol{R}'-\boldsymbol{R}(t)) \,, \text{ where} \begin{cases} \boldsymbol{R}(0)=\boldsymbol{R} \,, \\[2ex] \frac{d\boldsymbol(t)}{dt} = \frac{F(\boldsymbol{R}(t))}{2m} \,, \end{cases} \\[1ex]
&
G^{(3)}(\boldsymbol{R}',\boldsymbol{R},t) = \exp\big[-(E_{L}(\boldsymbol{R})-E_s)t \big] \, \delta(\boldsymbol{R}'-\boldsymbol{R}) \,.
\end{align}

Note herein that in the case of having more than one $\alpha$-channel, we propagate the quantity \cite{baena:2018}
\begin{equation}
{\cal F}(\boldsymbol{R},t) = \sum_\alpha f_{\alpha}(\boldsymbol{R},t) \,,
\end{equation}
such as
\begin{align}
{\cal F}(\boldsymbol{R}',t+\Delta t) &= \int dR \, \langle \boldsymbol{R}' | e^{-A \Delta t} | \boldsymbol{R} \rangle \, \sum_{\alpha'\alpha} e^{-V_{\alpha'\alpha}(\boldsymbol{R})\Delta t} \, f_\alpha(\boldsymbol{R},t) \nonumber \\ 
&
= \int dR \, \langle \boldsymbol{R}' | e^{-A \Delta t} | \boldsymbol{R} \rangle \, \omega(\boldsymbol{R},t) \, {\cal F}(\boldsymbol{R},t) \,,
\label{eq:TheF}
\end{align}
where the weight factor is introduced as 
\begin{equation}
\omega(\boldsymbol{R},t) = \frac{\sum_{\alpha'\alpha}e^{-V_{\alpha'\alpha}(\boldsymbol{R})\Delta t}f_{\alpha}(\boldsymbol{R},t)}{\sum_\alpha f_{\alpha}(\boldsymbol{R},t)} \,.
\label{eq:omega}
\end{equation}

The application of DMC method to hadron physics has been scarce. For an introduction of the DMC method see Ref. \cite{Kosztin:1996fh} and for applications to hadron physics see Refs. \cite{Bai:2016int,Gordillo:2020sgc, Gordillo:2021bra,Alcaraz-Pelegrina:2022fsi, Ma:2022vqf,Gordillo:2023tnz,Ma:2023int}.

\subsection{Wave function}
The total wave function of a tetraquark state can written as
\begin{equation}
\Psi = \vert space \rangle \otimes \vert spin \rangle \otimes \vert color \rangle \otimes \vert flavor \rangle.
\end{equation}
by taking into account possible color, spin and flavor configurations. The construction of the total wave function needs elaboration of Pauli principle. The possible spin wave functions for the $(q_1 q_2 \bar q_3 \bar q_4)$ tetraquark states with the $(q q \bar Q \bar Q)$ configuration are
\begin{eqnarray}
\chi_{1}=|(q_{1}q_{2})_{1}(\bar{Q}_{3}\bar{Q}_{4})_{1}\rangle_{2},\nonumber\\
\chi_{2}=|(q_{1}q_{2})_{1}(\bar{Q}_{3}\bar{Q}_{4})_{1}\rangle_{1},\nonumber\\
\chi_{3}=|(q_{1}q_{2})_{1}(\bar{Q}_{3}\bar{Q}_{4})_{1}\rangle_{0},\nonumber\\
\chi_{4}=|(q_{1}q_{2})_{1}(\bar{Q}_{3}\bar{Q}_{4})_{0}\rangle_{1},\nonumber\\
\chi_{5}=|(q_{1}q_{2})_{0}(\bar{Q}_{3}\bar{Q}_{4})_{1}\rangle_{1},\nonumber\\
\chi_{6}=|(q_{1}q_{2})_{0}(\bar{Q}_{3}\bar{Q}_{4})_{0}\rangle_{0},
\end{eqnarray}
with the spin alignments
\begin{equation}
\begin{aligned} \chi_{1}&=\uparrow\uparrow\uparrow\uparrow, \\
\chi_{2}&=\frac{1}{2}
\left(\uparrow\uparrow\uparrow\downarrow+\uparrow\uparrow\downarrow\uparrow-%
\uparrow\downarrow\uparrow\uparrow-\downarrow\uparrow\uparrow\uparrow%
\right), \\ \chi_{3}&=\frac{1}{\sqrt{3}}
\left(\uparrow\uparrow\downarrow\downarrow+\downarrow\downarrow\uparrow%
\uparrow\right) \\ &-\frac{1}{2\sqrt{3}}
\left(\uparrow\downarrow\uparrow\downarrow+\uparrow\downarrow\downarrow%
\uparrow+\downarrow\uparrow\uparrow\downarrow+\downarrow\uparrow\downarrow%
\uparrow\right) \\ \chi_{4}&=\frac{1}{\sqrt{2}}
\left(\uparrow\uparrow\uparrow\downarrow-\uparrow\uparrow\downarrow\uparrow%
\right), \\ \chi_{5}&=\frac{1}{\sqrt{2}}
\left(\uparrow\downarrow\uparrow\uparrow-\downarrow\uparrow\uparrow\uparrow%
\right), \\ \chi_{6}&=\frac{1}{2}
\left(\uparrow\downarrow\uparrow\downarrow-\uparrow\downarrow\downarrow%
\uparrow-\downarrow\uparrow\uparrow\downarrow+\downarrow\uparrow\downarrow%
\uparrow\right), \end{aligned}  \label{pspinT}
\end{equation}%
where subscript outside the paranthesis and bracket denotes the spin of the hadron.

The only color singlet configurations of  $q_1 q_2 \bar q_3 \bar q_4$ tetraquark states are $\textbf{6}_c \otimes \bar{\textbf{6}}_c$ or $\bar{\textbf{3}}_c \otimes \textbf{3}_c $ color configurations with the respective color wave functions
\begin{equation}
\phi _{1}=\left\vert {\left( q_{1}q_{2}\right) }^{6}{\left( \bar{q}_{3}%
\bar{q}_{4}\right) }^{\bar{6}}\right\rangle ,\quad \phi _{2}=\left\vert {%
\left( q_{1}q_{2}\right) }^{\bar{3}}{\left( \bar{q}_{3}\bar{q}_{4}\right) }%
^{3}\right\rangle. \label{colorT}
\end{equation}%

For the $q q \bar Q \bar Q$ configuration, we can write Eq. (\ref{colorT}) as

\begin{eqnarray}
\phi_{1}&=&|(q_{1}q_{2})^{6}(\bar{Q}_{3}\bar{Q}_{4})^{\bar{6}}\rangle\nonumber\\
        &=&\frac{1}{2\sqrt{6}}[2(rr\bar{r}\bar{r}+gg\bar{g}\bar{g}+bb\bar{b}\bar{b})+
rb\bar{b}\bar{r}+br\bar{b}\bar{r}\nonumber \\
        &&+gr\bar{g}\bar{r}+rg\bar{g}\bar{r}+gb\bar{b}\bar{g}+bg\bar{b}\bar{g}+
gr\bar{r}\bar{g}+rg\bar{r}\bar{g}\nonumber\\
        &&+gb\bar{g}\bar{b}+bg\bar{g}\bar{b}+rb\bar{r}\bar{b}+br\bar{r}\bar{b}]
\end{eqnarray}
and
\begin{eqnarray}
\phi_{2}&=&|(q_{1}q_{2})^{\bar{3}}(\bar{Q}_{3}\bar{Q}_{4})^{3}\rangle\nonumber\\
        &=&\frac{1}{2\sqrt{3}}(rb\bar{b}\bar{r}-br\bar{b}\bar{r}-gr\bar{g}\bar{r}+rg\bar{g}\bar{r}+
gb\bar{b}\bar{g}-bg\bar{b}\bar{g}\nonumber\\
        &&+gr\bar{r}\bar{g}-rg\bar{r}\bar{g}-gb\bar{g}\bar{b}+bg\bar{g}\bar{b}-rb\bar{r}\bar{b}+
br\bar{r}\bar{b}).
\end{eqnarray}

Concerning the color degree of freedom, the heavy quark is treated as an $SU(3)$ singlet. 

Armed with these considerations, we give the wave function bases in the notation of
$|(q_1q_2)^{\rm color}_{\rm spin}(\bar{Q}_3\bar{Q}_4)^{\rm color }_{\rm spin }\rangle_{\text{total spin}}$:

\begin{eqnarray}\label{basis-vectors}
&&\phi_1\chi_1=|(q_1q_2)^{6}_1(\bar{Q}_3\bar{Q}_4)^{\bar{6}}_1\rangle_2\delta_{12}^S\delta_{34},\nonumber\\
&&\phi_2\chi_1=|(q_1q_2)^{\bar{3}}_1(\bar{Q}_3\bar{Q}_4)^{3}_1\rangle_2\delta_{12}^A,\nonumber\\
&&\phi_1\chi_2=|(q_1q_2)^{6}_1(\bar{Q}_3\bar{Q}_4)^{\bar{6}}_1\rangle_1\delta_{12}^S\delta_{34},\nonumber\\
&&\phi_2\chi_2=|(q_1q_2)^{\bar{3}}_1(\bar{Q}_3\bar{Q}_4)^{3}_1\rangle_1\delta_{12}^A,\nonumber\\
&&\phi_1\chi_3=|(q_1q_2)^{6}_1(\bar{Q}_3\bar{Q}_4)^{\bar{6}}_1\rangle_0\delta_{12}^S\delta_{34},\nonumber\\
&&\phi_2\chi_3=|(q_1q_2)^{\bar{3}}_1(\bar{Q}_3\bar{Q}_4)^{3}_1\rangle_0\delta_{12}^A,\\
&&\phi_1\chi_4=|(q_1q_2)^{6}_1(\bar{Q}_3\bar{Q}_4)^{\bar{6}}_0\rangle_1\delta_{12}^S,\nonumber\\
&&\phi_2\chi_4=|(q_1q_2)^{\bar{3}}_1(\bar{Q}_3\bar{Q}_4)^{3}_0\rangle_1\delta_{12}^A\delta_{34},\nonumber\\
&&\phi_1\chi_5=|(q_1q_2)^{6}_0(\bar{Q}_3\bar{Q}_4)^{\bar{6}}_1\rangle_1\delta_{12}^A\delta_{34},\nonumber\\
&&\phi_2\chi_5=|(q_1q_2)^{\bar{3}}_0(\bar{Q}_3\bar{Q}_4)^{3}_1\rangle_1\delta_{12}^S,\nonumber\\
&&\phi_1\chi_6=|(q_1q_2)^{6}_0(\bar{Q}_3\bar{Q}_4)^{\bar{6}}_0\rangle_0\delta_{12}^A,\nonumber\\
&&\phi_2\chi_6=|(q_1q_2)^{\bar{3}}_0(\bar{Q}_3\bar{Q}_4)^{3}_0\rangle_0\delta_{12}^S\delta_{34}\nonumber. \label{colorspinT}
\end{eqnarray}
One should be aware of the wave functions which are not allowed for a given quantum numbers. It can be seen in Eq. (\ref{colorspinT}) that, $\delta_{12}^S$, $\delta_{12}^A$, and $\delta_{34}$ factors are introduced to reflect flavor symmetry. If light quarks in $(q_1 q_2)$ are symmetric then $\delta_{12}^S=0$. Accordingly if light quarks in $(q_1 q_2)$ are antisymmetric then $\delta_{12}^A=0$. If the two quarks are heavy and identical, then we have $\delta_{34}=0$. Apart from these considerations, when the factor cannot be equal to 0, one can use its value as 1. A similar logic was used in Ref. \cite{Zhang:2021yul}.

\subsection{Magnetic Moment}
The magnetic moment of a hadron arises from its charge distribution and angular momentum distribution. Not only it presents valuable information about the internal structure of the hadron but also the shape of the hadron. Furthermore, magnetic moment is an excellent probe for quark-gluon dynamics inside the hadron since it is the leading-order response of a bound state to a weak external magnetic moment \cite{Ozdem:2023rkx}. 

The magnetic moment of a doubly-heavy tetraquark system can be written as 
\begin{equation}
\mu=\langle \phi_{IJ} \vert \hat{\mu}_{m} \vert \phi_{IJ} \rangle,
\end{equation}
where $\hat{\mu}_{m}=\sum_{i=1}^4\frac{\hat{Q}_i}{2m_i}\hat{\sigma}_i^z$ is the magnetic moment operator, $\hat{Q}_i$ is the electric charge operator of the $i$-th
quark, $\sigma^z_i$ is the third component ($z$-component) of Pauli matrix and $\phi_{IJ}$ is the eigenvector of those states.

\section{\label{sec:level3}Numerical Results and Discussion}
The DMC algorithm can be carried out to obtain the desired solution to Eq.~(\ref{eq:TheF}) equation with the formalism described in Ref. \cite{Gordillo:2020sgc}. We follow the same algorithm.

The main idea can be described as follows: The kinetic and potential energy terms of the Schrödinger equation correspond to the diffusion term (source/sink) in the diffusion equation
\begin{equation}
\frac{\partial u(\boldsymbol{r},t)}{\partial t}=\nabla \cdot \left[ D(u(\boldsymbol{r},t),\boldsymbol{r}) \nabla u(\boldsymbol{r},t) \right]
\end{equation}
where $u(\boldsymbol{r},t)$ is the density of the diffusing matter and $D(u(\boldsymbol{r},t),\boldsymbol{r})$ is the diffusion term or coefficient.
The Schrödinger equation with imaginary time is a kind of diffusion equation with source/sink mechanism. It can be solved by simulating random walks which are subject to birth/death processes resulting from the source/sink term. In the DMC algorithm, the wave function is represented by random walks of many particles. Each walker will walk randomly in finite small time steps and experience death or birth. After sufficiently long time, the diffusion of the walkers will approach the ground state wave function.

There are two theoretical errors in this model which have different sources. The first one is statistical related to DMC method itself and the second one is related to quark model parameters. We utilize Monte Carlo simulations comprising 50000 steps. We use $1 \times 10^4$ walkers in the simulations to implement the movement of each quark. Any implementation of DMC method leads to a point where the distribution of the walkers should be uncorrelated. To avoid this situation, we consider in the averages the values for every 500 steps and we drop the first $10^5$ values to prevent any effect of  chosen initial configuration. The resulting energy is averaged over the last 10000 steps with a standard deviation $\simeq 1\,\text{MeV}$.

Before we present our results, we give possible corresponding thresholds in Table \ref{threshold} regarding the states that are studied. We also present our results according to the strangeness quantum number. 

\begin{table}[H]
    \centering
    \caption{Corresponding thresholds for the doubly heavy tetraquark systems of this work.}
    \label{threshold}
\begin{tabular*}{\hsize}{@{}@{\extracolsep{\fill}}cccccc@{}}
\hline 
\hline 
\multirow{2}{*}{System} & \multirow{2}{*}{$I$} & \multicolumn{3}{c}{Thresholds}\tabularnewline
\cline{3-5} 
 &  & $J^P=0^{+}$ & $J^P=1^{+}$ & $J^P=2^{+}$\tabularnewline
\hline 
\multirow{2}{*}{$nn\bar{c}\bar{c}$} & 0 &  & $DD^{*}$ & \tabularnewline
 & 1 & $DD$ & $DD^{*}$ & $D^{*}D^{*}$\tabularnewline
$ss\bar{c}\bar{c}$ & 0 & $D_{s}D_{s}$ & $D_{s}D_{s}^{*}$ & $D_{s}^{*}D_{s}^{*}$\tabularnewline
\hline 
\multirow{2}{*}{$nn\bar{b}\bar{b}$} & 0 & & $\bar{B}\bar{B}^{*}$ & \tabularnewline
 & 1 & $\bar{B}\bar{B}$ & $\bar{B}\bar{B}^{*}$ & $\bar{B}^{*}\bar{B}^{*}$\tabularnewline
$ns\bar{b}\bar{b}$ & $\frac{1}{2}$ & $\bar{B}\bar{B}_{s}$ & $\bar{B}_{s}^{*}\bar{B}/\bar{B}^{*}\bar{B}_{s}$ & $\bar{B}^{*}\bar{B}_{s}^{*}$\tabularnewline
$ss\bar{b}\bar{b}$ & 0 & $\bar{B}_{s}\bar{B}_{s}$ & $\bar{B}_{s}\bar{B}_{s}^{*}$ & $\bar{B}_{s}^{*}\bar{B}_{s}^{*}$\tabularnewline
\hline 
\multirow{2}{*}{$nn\bar{c}\bar{b}$} & 0 & $D\bar{B}$ & $D^{*}\bar{B}/D\bar{B}^{*}$ & $D^{*}\bar{B}^{*}$\tabularnewline
 & 1 & $D\bar{B}$ & $D^{*}\bar{B}/D\bar{B}^{*}$ & $D^{*}\bar{B}^{*}$\tabularnewline
$ss\bar{c}\bar{b}$ & 0 & $\bar{B}_{s}D_{s}$ & $D_{s}^{*}\bar{B}_{s}/D_{s}\bar{B}_{s}^{*}$ & $\bar{B}_{s}^{*}D_{s}^{*}$\tabularnewline
\hline 
\hline
\end{tabular*}
\end{table}

\subsection{Tetraquarks with Strangeness-Zero (S=0)}
The quark content for these tetraquark states are $nn\bar c \bar c$, $nn\bar b \bar b$, and $nn\bar c \bar b$. In Table \ref{tab:2heavytetraquark1}, our results for the mass and magnetic moment values are listed. 

\renewcommand{\tabcolsep}{0.33cm} \renewcommand{\arraystretch}{1.1}
\begin{table*}[H]
\centering
\caption{Mass spectrum and magnetic moments of doubly heavy $nn\bar{c}\bar{c}$, $nn\bar{b}\bar{b}$ and $nn\bar{c}\bar{b}$ tetraquarks, where $n=u ~ \text{or} ~ d$. Mass results are in units of MeV. Magnetic moments (in units of nuclear magneton $\protect\mu_{N}$, $\mu_p=2.79285\mu_N$) are organized in the order of $I_{3}=1,\,0,\,-1$ for $I=1$.}
\label{tab:2heavytetraquark1}%
\begin{tabular}{cccccc}
\hline\hline
\textrm{State} & $J^{P}$ &   & Mass & $%
\mu$ \\ \hline
\multirow{2}{*}{${(nn\bar{c}\bar{c})}^{I=1}$} & \multirow{2}{*}{$0^{+}$} &
&  4283 & -  \\
&  &  &  4062  & -  \\
& $1^{+}$ &  &   4104  & 1.14, -0.07, 1.26  \\
& $2^{+}$ &  & 4207  & 2.66, -0.06, 2.39  \\
\hline
\multirow{2}{*}{${(nn\bar{c}\bar{c})}^{I=0}$} & \multirow{2}{*}{$1^{+}$} &
 &   3892 & -0.28 &  \\
&  &  &  4130 & 0.35 &  \\
\hline
\multirow{2}{*}{${(nn\bar{b}\bar{b})}^{I=1}$} & \multirow{2}{*}{$0^{+}$} &
&   10748 & -  \\
&  &  &  10624 & -  \\
& $1^{+}$ &  &   10680 & -1.02, 0.28, -0.64  \\
& $2^{+}$ &  &  10702 & -2.32, 0.47, -1.28 \\ 
\hline
\multirow{2}{*}{${(nn\bar{b}\bar{b})}^{I=0}$} & \multirow{2}{*}{$1^{+}$} &
 &   10338 & 0.32 & \\
&  &  &   10624 & 0.69 &  \\
\hline
\multirow{2}{*}{${(nn\bar{c}\bar{b})}^{I=1}$} & \multirow{2}{*}{$0^{+}$} &
 &  7620 & -  \\
&  &  &  7396 & -  \\
& \multirow{3}{*}{$1^{+}$} &  &   7510 & 1.84, 0.12, -1.36  \\
&  &  &  7448 & 0.09, 0.12, 0.21 &  \\
&  &  & 7542 & 1.94, 0.41, -0.59 &  \\
& $2^{+}$ &  &   7568 & 2.47, 0.32, -1.68  \\ 
\hline
\multirow{2}{*}{${(nn\bar{c}\bar{b})}^{I=0}$} & \multirow{2}{*}{$0^{+}$} &
 &   7108 & -  \\
&  & &   7201 & - \\
& \multirow{3}{*}{$1^{+}$} &  & 7084  & 0.74
\\
&  &  &   7324 & 0.70  \\
&  &  &   7204 & 0.45  \\
& $2^{+}$ &  &   7360 & -0.54 \\ \hline\hline
\end{tabular}%
\end{table*}

\begin{widetext}

\begin{figure}[!htb]
\begin{center}
\includegraphics[totalheight=8cm,width=8cm]{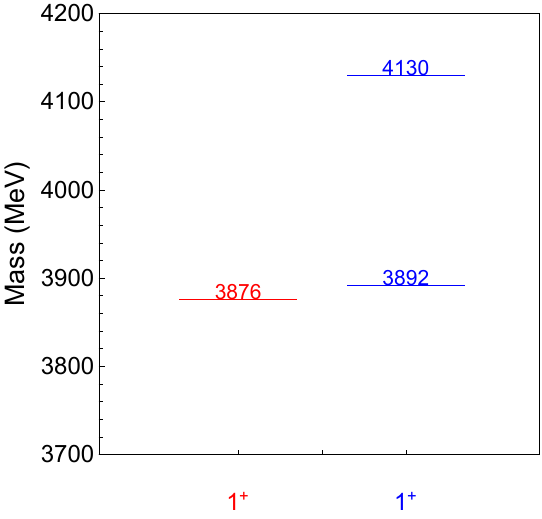}
\includegraphics[totalheight=8cm,width=8cm]{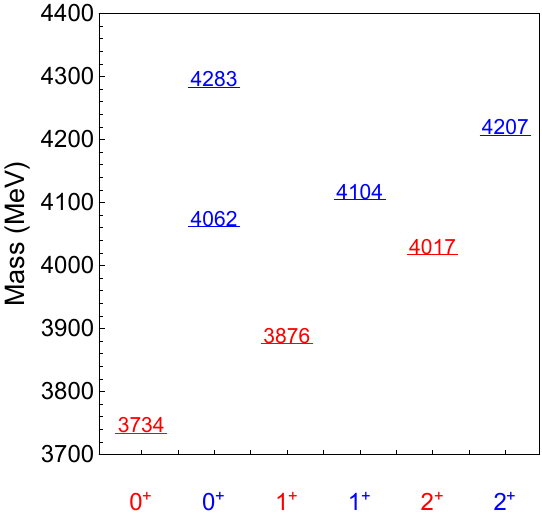}
\end{center}
\caption{Mass spectrum of $nn\bar c \bar c$ states with $I=0$ (left), $I=1$ (right) and corresponding $J^P$ values. Red color refers to threshold values whereas blue one refers to obtained results.}
\label{fig:nncc}
\end{figure}

\end{widetext}

It can be seen from the table that, the mass splittings in the  
$nn\bar{c}\bar{c}$ system are 221 MeV for $I=1$ and 238 MeV for $I=0$. In the $nn\bar{b}\bar{b}$ system, for $I=1$ the mass splitting is 124 MeV and for $I=0$ it is 286 MeV. For the mixed heavy quark system, $nn\bar{c}\bar{b}$, the mass splittings are 224 MeV for $I=1$ and 276 MeV. One can also compare mass spectra with the corresponding thresholds which is shown in Fig. \ref{fig:nncc}. 

 We can see from the figure that obtained mass values for $I=1$ case are above corresponding thresholds. They should be resonance states. A special attention should be made on the $I=0$ states. In the case of $I=0$, one of the  $J^P=1^+$ states is high above than the $DD^\ast$ threshold. The other is just a few MeV above. The mass and magnetic moment of the $T_{cc}^+$ are predicted to be $M=3892 ~ \text{MeV}$ and $\mu=0.28\mu_N$, respectively. The obtained mass agree well with the observed $T_{cc}^+$ mass, $M=3875$ MeV. Ref. \cite{Deng:2021gnb} calculated $I(J^P)=0(1^+)$ $T_{cc}^+$ magnetic moment as $\mu=0.18\mu_N$ in compact configuration and $\mu=0.13 \mu_N$ in deuteron-like configuration. In Ref. \cite{Azizi:2021aib}, the magnetic moments of $T_{cc}^+$ state are obtained in both diquark-antidiquark and molecular configurations via light-cone QCD sum rule formalism. They obtained magnetic moment as $\mu=0.66^{+0.34}_{-0.23}\mu_N$ with diquark-antidiquark configuration and $\mu=0.43^{+0.23}_{-0.22}\mu_N$ with molecular configuration.  Ref. \cite{Zhang:2021yul} obtained magnetic moment of $T_{cc}^+$ as $\mu=0.88 \mu_N$ by using MIT bag model. Ref. \cite{Lei:2023ttd} calculated magnetic moment as $\mu=-0.09\mu_N$  by using constituent quark model. Ref. \cite{Wu:2022gie} obtained a magnetic moment of  $\mu=0.732\mu_N$ by using heavy antiquark-diquark symmetry in the constituent quark model. The calculated results for the magnetic moment of $T_{cc}^+$ in general are not consistent with each other. 

\begin{widetext}

\begin{figure}[!htb]
\begin{center}
\includegraphics[totalheight=8cm,width=8cm]{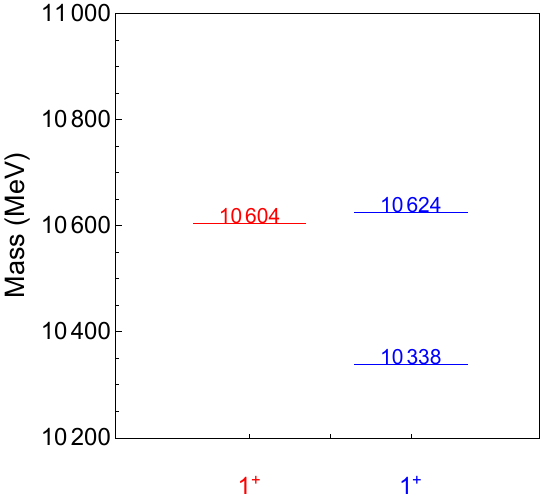}
\includegraphics[totalheight=8cm,width=8cm]{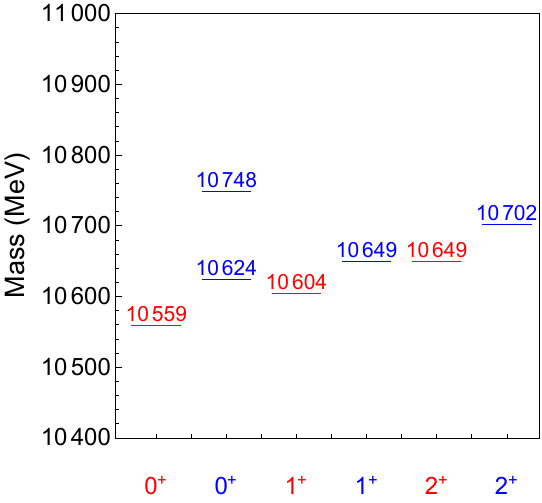}
\end{center}
\caption{Mass spectrum of $nn\bar b \bar b$ states with $I=0$ (left), $I=1$ (right) with corresponding $J^P$ values. Red color refers to threshold values whereas blue one refers to obtained results.}
\label{fig:nnbb}
\end{figure}

\end{widetext}

In Figure \ref{fig:nnbb}, mass spectrum of $nn\bar b \bar b$ states can be seen. The results of $I=1$ case are above corresponding thresholds. They should be resonance states. In the case of $I=0$, the state which has a mass of 10624 MeV is 20 MeV above $\bar{B} \bar{B}^\ast$ threshold whereas the other is 266 MeV below. The obtained mass of this state in our model is $M=10338 ~\text{MeV}$. This should be a deeply bound state. Considering magnetic moment of this state which is named as $T_{bb}^-$, we obtain $\mu=-0.32\mu_N$.  Ref. \cite{Deng:2021gnb} calculated $I(J^P)=0(1^+)$ $T_{bb}^-$ magnetic moment as $\mu=0.64~\mu_N$ in compact configuration and as $\mu=0.49 \mu_N$ in deuteron-like configuration. In Ref. \cite{Ozdem:2022yhi}, magnetic moments of $T_{bb}$ states with $J^P=1^+$ are obtained in light-cone QCD sum rule using molecular picture. The results follow as $\mu=(1.72 \pm 0.67)\mu_N$ for $B^- B^{*-}$, $\mu=(1.38 \pm 0.56) \mu_N$ for $B^0 B^{*-}$, $\mu= (-0.44 \pm 0.17)\mu_N$ for $B^- B^{*0} $, and $ \mu=(-0.77 \pm 0.28)\mu_N$ for $B^0 B^{*0}$. Ref. \cite{Zhang:2021yul} obtained magnetic moment of $T_{bb}^-$ as $\mu=0.18 \mu_N$ whereas Ref. \cite{Wu:2022gie} obtained  $\mu=-0.124\mu_N$ for $T_{bb}^-$.  As in the $T_{cc}^+$ case, the calculated results for the magnetic moment of $T_{bb}^-$ in general are not consistent with each other. 

\begin{widetext}

\begin{figure}[!htb]
\begin{center}
\includegraphics[totalheight=8cm,width=8cm]{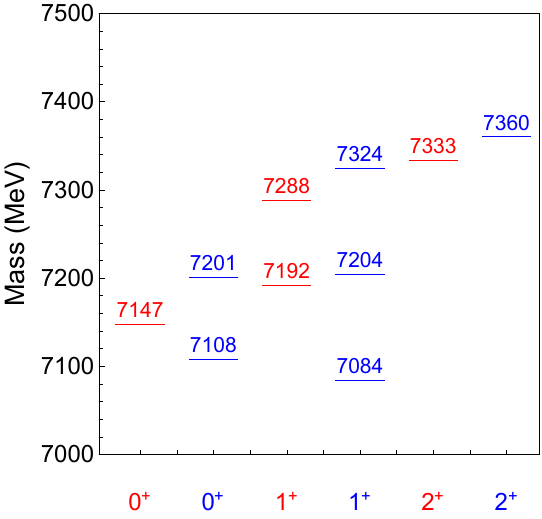}
\includegraphics[totalheight=8cm,width=8cm]{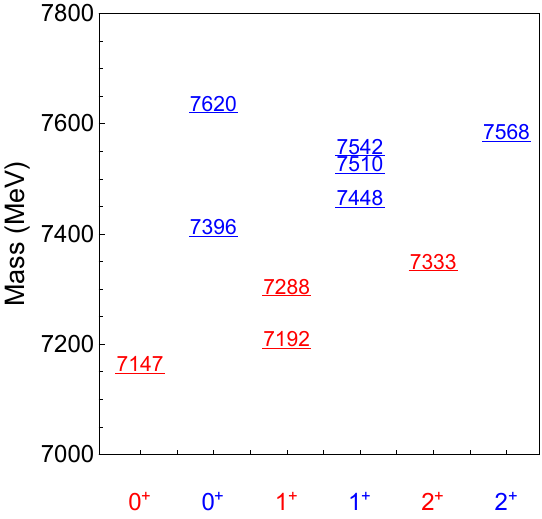}
\end{center}
\caption{Mass spectrum of $nn\bar c \bar b$ states with $I=0$ (left), $I=1$ (right) with corresponding $J^P$ values. Red color refers to threshold values whereas blue one refers to obtained results.}
\label{fig:nncb}
\end{figure}

\end{widetext}

In Figure \ref{fig:nncb}, mass spectrum of $nn\bar c \bar b$ states can be seen. All the results corresponding to $I=1$ are above corresponding thresholds. These states should be resonances. We have bound state candidates in the $I=0$ channel. The $J^P=0^+$ state which has a mass of 7108 MeV is 39 MeV below $D\bar{B}$ threshold. In $J^P=1^+$ states, the state which has a mass of 7084 is 108 MeV below $D \bar{B}^\ast$ threshold and 204 MeV below $D^\ast \bar{B}$ threshold. The state having a mass of 7204 MeV is 84 MeV below $D^\ast \bar{B}$ threshold. We also obtain magnetic moments of $I=0$ and $I=1$ $nn\bar c \bar b$ states. Focusing on $I=0$ states, we obtain 
$\mu=0.74 \mu_N$ and $\mu=0.45 \mu_N$ for two low-lying $J^P=1^+$ states. Ref. \cite{Deng:2021gnb} obtained $I(J^P)=0(1^+)$ $T_{bc}$ magnetic moment as $\mu=0.66 \mu_N$ in compact configuration and as $\mu=0.59 \mu_N$ in deuteron-like configuration. Our results agree with these results. 

\subsection{Tetraquarks with Strangeness-One (S=-1)}
The quark content for these tetraquark states are $ns\bar c \bar c$ and $ns\bar b \bar b$. We present mass spectrum and magnetic moment results in 
Table \ref{tab:2heavytetraquark2}. As can be seen from table, in the $ns\bar c \bar c$ system the mass splitting is 285 MeV for $J^P=0^+$ states and 145 MeV for $J^P=1^+$ states. In the $ns\bar b \bar b$ system, the mass splitting is 131 MeV for $J^P=0^+$ states and 94 MeV for $J^P=1^+$ states.

\renewcommand{\tabcolsep}{0.52cm} \renewcommand{\arraystretch}{1.1}
\begin{table*}[!htb]
\centering
\setlength{\abovecaptionskip}{0.4cm}
\caption{Mass spectrum and magnetic moments of doubly heavy $ns\bar{c}\bar{c}$ and $ns\bar{b}\bar{b}$ tetraquarks, where $n=u ~ \text{or} ~ d$. Mass results are in units of MeV. Magnetic moments (in units of nuclear magneton $\protect\mu_{N}$, $\mu_p=2.79285\mu_N$) are organized in the order of $I_{3}=1/2,\,-1/2$ for $I=1$. }
\label{tab:2heavytetraquark2}%
\begin{tabular}{cccc}
\hline\hline
\textrm{State} & $J^{P}$ &  Mass & $\mu$ \\ \hline
\multirow{2}{*}{$ns\bar{c}\bar{c}$} & \multirow{2}{*}{$0^{+}$} & 4349 & - \\
 &  & 4064 & - \\
& \multirow{3}{*}{$1^{+}$} & 4056& 0.43, -0.67\\
&   & 4201 & 1.24, -1.96  \\
&   & 4128 & 0.52, -0.75  \\
& $2^{+}$ & 4314 & 0.24, -2.76  \\\hline
\multirow{2}{*}{$ns\bar{b}\bar{b}$} & \multirow{2}{*}{$0^{+}$}  & 10852 & -  \\
&  & 10721 & - \\
& \multirow{3}{*}{$1^{+}$} &  10736 & 0.48,
-0.66  \\
&   & 10778 & 1.11, -1.35 \\
&  & 10684 & -0.09, -0.12  \\
& $2^{+}$ & 10747 & 1.36, -1.98  \\ \hline\hline
\end{tabular}%
\end{table*}

\begin{widetext}

\begin{figure}[!htb]
\begin{center}
\includegraphics[totalheight=8cm,width=8cm]{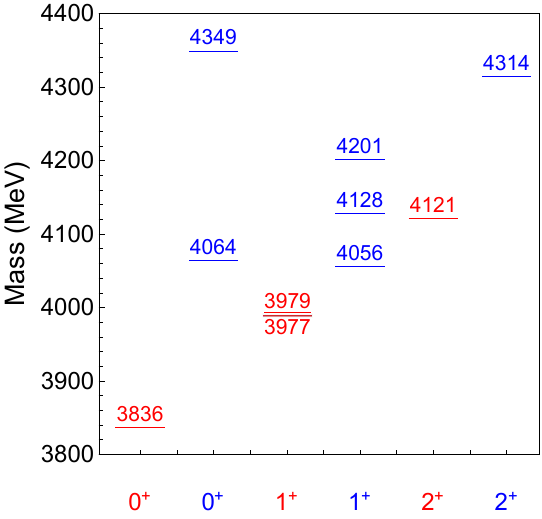}
\includegraphics[totalheight=8cm,width=8cm]{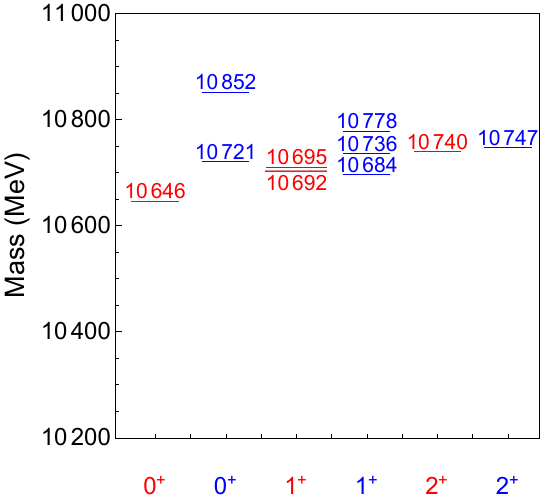}
\end{center}
\caption{Mass spectrum of $ns\bar c \bar c$ (left) and  $ns\bar{b}\bar{b}$ (right)  tetraquarks with corresponding $J^P$ values. Red color refers to threshold values whereas blue one refers to obtained results.}
\label{fig:nsccbb}
\end{figure}

\end{widetext}

In Figure \ref{fig:nsccbb}, mass spectrum of $ns\bar{c}\bar{c}$ and $ns\bar{b}\bar{b}$ tetraquarks can be seen. In the $ns\bar{c}\bar{c}$ systems, all obtained mass values are above corresponding thresholds. They should be resonance states. In the $ns\bar{b}\bar{b}$ systems,  the $J^P=1^+$ state with a mass of $M=10684$ is below corresponding thresholds $\bar{B}_s^\ast \bar{B}/\bar{B}^\ast \bar{B}_s$. This could be a bound state candidate. The $J^P=2^+$ state is just 7 MeV above corresponding $\bar{B}^\ast \bar{B}_s^\ast$ state. This state seems to be a resonance state but also could be a shallow bound state candidate. Focusing on this state, in Ref. \cite{Azizi:2023gzv}, $J^P=1^+$ $T_{QQ\bar q\bar s}$ and $T_{QQ\bar s\bar s}$ states are studied as diquark-antidiquark tetraquarks and magnetic moments are calculated with using heavy axial vector diquark-light scalar antidiquark type current denoted as $J^1$, heavy scalar diquark-light axial vector antidiquark type current denoted as $J^2$, heavy tensor diquark-light axial vector antidiquark type current denoted as $J^3$ and heavy vector diquark-light pseudotensor antidiquark type current denoted as $J^4$. According to their work, $J^1$ gives $\mu=(-0.46 \pm 0.07) \mu_N$, $J^2$ gives $\mu=(-0.34 \pm 0.12) \mu_N$, $J^3$ gives $\mu=(-2.37 \pm 0.50) \mu_N$ and $J^4$ gives $\mu=(-2.90 \pm 0.36) \mu_N$ for  $T_{bb\bar u \bar s}$ state. For $T_{bb\bar d \bar s}$ state, $J^1$ gives $\mu=(-0.59 \pm 0.09) \mu_N$, $J^2$ gives $\mu=(-0.68 \pm 0.24) \mu_N$, $J^3$ gives $\mu=(-1.37 \pm 0.17) \mu_N$ and $J^4$ gives $\mu=(-1.45 \pm 0.33) \mu_N$. Our results are compatible with this work. In the same study considering $T_{cc\bar u \bar s}$ state, $J^1$ gives $\mu=(0.22 \pm 0.07) \mu_N$, $J^2$ gives $\mu=(0.18 \pm 0.06) \mu_N$, $J^3$ gives $\mu=(1.09 \pm 0.36) \mu_N$ and $J^4$ gives $\mu=(1.04 \pm 0.35) \mu_N$. For $T_{cc\bar d \bar s}$ state, $J^1$ gives $\mu=(0.15 \pm 0.05) \mu_N$, $J^2$ gives $\mu=(0.36 \pm 0.11) \mu_N$, $J^3$ gives $\mu=(1.70 \pm 0.57) \mu_N$ and $J^4$ gives $\mu=(2.08 \pm 0.64) \mu_N$. Our results are compatible with this work. 

\subsection{Tetraquarks with Strangeness-Two (S=-2)}
The quark content for these tetraquark states are $ss\bar c \bar c$ and $ss\bar b \bar b$. We present mass spectrum and magnetic moment results in Table \ref{tab:2heavytetraquark3}. As can be seen from table, in the $ss\bar c \bar c$ system the mass splitting are 269 MeV for $J^P=0^+$ states, 104 MeV for $ss\bar b \bar b$ $J^P=0^+$ states. In the $ss\bar c \bar b$ systems, the mass splittings are 180 MeV for $J^P=0^+$ states and 105 MeV for $J^P=1^+$ states.

\renewcommand{\tabcolsep}{0.60cm} \renewcommand{\arraystretch}{1.1}
\begin{table}[!htb]
\centering
\caption{Mass spectrum and magnetic moments of doubly heavy $ns\bar{c}\bar{c}$ and $ns\bar{b}\bar{b}$ tetraquarks, where $n=u ~ \text{or} ~ d$. Mass results are in units of MeV. Magnetic moments are presented in units of nuclear magneton $\protect\mu_{N}$, $\mu_p=2.79285\mu_N$. }
\label{tab:2heavytetraquark3}%
\begin{tabular}{ccccc}
\hline\hline
\textrm{State} & $J^{P}$ &  Mass & $%
\mu$  \\ \hline
\multirow{2}{*}{$ss\bar{c}\bar{c}$} & \multirow{2}{*}{$0^{+}$} &  4474 & -  \\
 &  & 4205 & - &\\
& $1^{+}$ &   4323 & -1.33  \\
& $2^{+}$ &  4381 & -1.76 \\\hline
\multirow{2}{*}{$ss\bar{b}\bar{b}$} & \multirow{2}{*}{$0^{+}$} &  10972 & -  \\
&  & 10868& -  \\
& $1^{+}$ & 10908 & 0.73  \\
& $2^{+}$ & 10926 & 1.42  \\ \hline
\multirow{2}{*}{$ss\bar{c}\bar{b}$} & \multirow{2}{*}{$0^{+}$} &   7833 & -  \\
&  & 7653 & -  \\
& \multirow{3}{*}{$1^{+}$} & 7764 & -0.86
\\
&  &   7816& -0.35  \\
&  &  7711 & -0.21  \\
& $2^{+}$ &  7862& -2.09 \\ \hline\hline
\end{tabular}%
\end{table}

\begin{figure}[!htb]
\begin{center}
\includegraphics[totalheight=8cm,width=8cm]{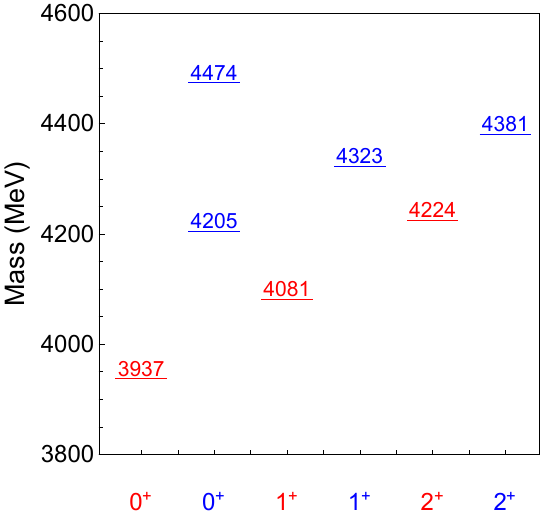}
\end{center}
\caption{Mass spectrum of $ss\bar c \bar c$ tetraquark states. Red color refers to threshold values whereas blue one refers to obtained results.}
\label{fig:sscc}
\end{figure}

\begin{figure}[!htb]
\begin{center}
\includegraphics[totalheight=8cm,width=8cm]{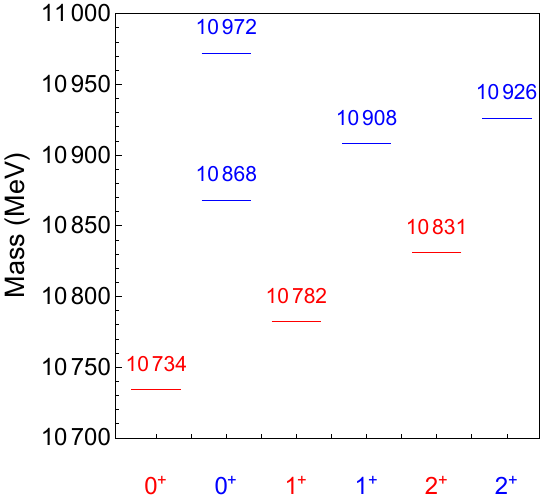}
\end{center}
\caption{Mass spectrum of $ss\bar b \bar b$ tetraquark states. Red color refers to threshold values whereas blue one refers to obtained results.}
\label{fig:ssbb}
\end{figure}

\begin{figure}[!htb]
\begin{center}
\includegraphics[totalheight=8cm,width=8cm]{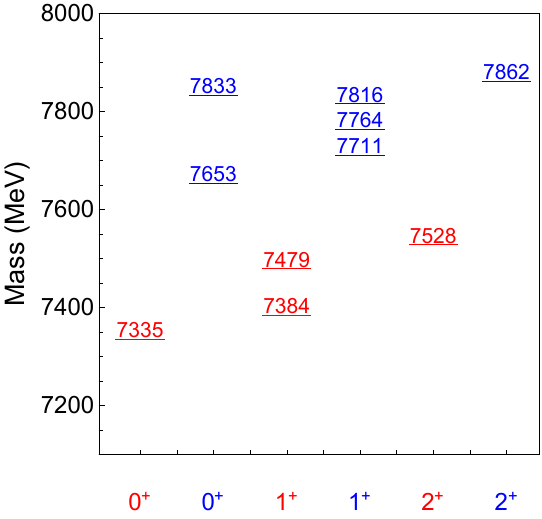}
\end{center}
\caption{Mass spectrum of $ss\bar c \bar b$ tetraquark states. Red color refers to threshold values whereas blue one refers to obtained results.}
\label{fig:sscb}
\end{figure}

In Figures \ref{fig:sscc}, \ref{fig:ssbb}, and \ref{fig:sscb} mass values of $ss\bar c \bar c$, $ss \bar b \bar b$, and $ss \bar c \bar b$ states are located. As can be seen from figures all the states are above their corresponding thresholds, therefore they should be resonance states. Considering magnetic moments, Ref. \cite{Azizi:2023gzv} calculated magnetic moments of $T_{cc \bar s \bar s}$ and $T_{bb \bar s \bar s}$ states in light cone QCD sum rule formalism. The results are $\mu=(1.52 \pm 0.46) \mu_N$ for $J^3$ type current and $\mu=(1.90 \pm 0.56) \mu_N$ for $J^4$ type current. Our magnetic moment results of $ss \bar c \bar c$ are compatible with these results. For the $T_{bb \bar s \bar s}$ state, they obtained $\mu=(-2.90 \pm 0.23) \mu_N$ for $J^3$ type current and $\mu=(-3.20 \pm 0.48) \mu_N$ for $J^4$ type current. Our results are not as compatible as in the previous case.

For the wave functions in this work, the general ground state can be the mixing of $\phi _{2} \chi_{3}$ and $\phi _{1}\chi _{6}$ for the isotriplet tetraquarks with $J^{P}=0^{+}$. In the case of isosinglet tetraquarks with $J^{P}=1^{+}$, the wave function can be mixing of $\phi_{2}\chi _{5}$ and $\phi _{1}\chi _{4}$ for  $nn\bar{Q}\bar{Q}$ tetraquarks and mixing of $\phi _{2}\chi _{2}$, $\phi _{2}\chi _{5}$, $\phi
_{1}\chi _{4}$ for  $nn\bar{Q}\bar{Q}$ tetraquarks. In the $nn \bar{c} \bar{b}$ tetraquarks, for $I(J^P)=1(0^+)$ the wave function can be mixing of $\phi_{2}\chi _{3}$ and $\phi _{1}\chi _{6}$, for $I(J^P)=0(0^+)$ the wave function can be mixing of $\phi_{1}\chi _{3}$ and $\phi _{2}\chi _{6}$, for $I(J^P)=1(1^+)$ the wave function can be mixing of $\phi_{2}\chi _{2}$, $\phi _{2}\chi _{4}$, $\phi _{1}\chi _{5}$, and for $I(J^P)=0(1^+)$ the wave function can be mixing of $\phi_{2}\chi _{2}$, $\phi _{2}\chi _{5}$, $\phi _{1}\chi _{4}$.

We compare our results with some of the works that studied same states. In Table \ref{tab:comparisonmass} we present our results with some other works conducted in different frameworks. As can be seen from table, our results are in agreement with the cited works. We also compare our mass results in Table \ref{tab:comparisonDMC} with a recent work conducted in DMC \cite{Ma:2023int}. The agreement of the results can be seen.

\renewcommand{\tabcolsep}{0.26cm} \renewcommand{\arraystretch}{1.1}
\begin{table*}[!htb]
\centering
\caption{Comparison of calculated mass (in MeV) with different calculations conducted in different models for double heavy tetraquarks. The masses before and after slash stand for
color states splitting. Refs. \protect\cite{Zhang:2021yul}, \protect\cite{Luo:2017eub}, \protect\cite{Lu:2020rog} employ the CMI model and Ref. \protect\cite{Ebert:2007rn} employ relativistic quark model.}
\label{tab:comparisonmass}%
\begin{tabular}{ccccccc}
\hline\hline
\textrm{State} & $J$ & \textrm{This\ work} & \cite%
{Luo:2017eub} & \cite{Zhang:2021yul}& \cite{Lu:2020rog} &
\cite{Ebert:2007rn} \\ \hline
${(nn\bar{c}\bar{c})}^{I=1}$ & 0 & 4062/4283 & 4078/4356 & 4032/4342 &
4195/4414 & 4056 \\
& 1 & 4104 & 4201 & 4117 & 4268 & 4079 \\
& 2 & 4207 & 4271 & 4179 & 4318 & 4118 \\
${(nn\bar{c}\bar{c})}^{I=0}$ & 1 & 3892/4130 & 4007/4204 & 3925/4205 &
4041/4313 & 3935 \\
${(nn\bar{b}\bar{b})}^{I=1}$ & 0 & 10624/10748 & 10841/10937 &
10834/11092 & 10765/11019 & 10648 \\
& 1 & 10680 & 10875 & 10854  & 10779 & 10657 \\
& 2 & 10702 & 10897 & 10878 & 10799 & 10673 \\
${(nn\bar{b}\bar{b})}^{I=0}$ & 1 & 10338/10624 & 10686/10821 &
10654/10982  & 10550/10951 & 10502 \\
${(nn\bar{c}\bar{b})}^{I=1}$ & 0 & 7396/7620 & 7457/7643 & 7438/7714  &
7519/7740 & 7383 \\
& \multirow{2}{*}{1} & 7510/7542 & 7473/7548 & 7465/7509 &
7537/7561 & \multirow{2}{*}{7396/7403} \\
& & 7448 & 7609 & 7699  & 7729 & \\
& 2 & 7568 & 7582 & 7531 & 7586 & 7422 \\
${(nn\bar{c}\bar{b})}^{I=0}$ & 0 & 7108/7201 & 7256/7429 & 7260/7502 &
7297/7580 & 7239 \\
& \multirow{2}{*}{1} & 7084/7204 & 7321/7431 & 7288/7518 &
7.325/7.607 & \multirow{2}{*}{7246} \\
& & 7324 & 7516 & 7605 & 7666 & \\
& 2 & 7360 & 7530 & 7483 & 7697 & - \\
$ns\bar{c}\bar{c}$ & 0 & 4165/4349 & 4236/4514 & 4165/4429 &
4323/4512 & 4221 \\
& \multirow{2}{*}{1} & 4056/4128& 4225/4363 & 4091/4247 &
4232/4394 & \multirow{2}{*}{4143/4239} \\
& & 4201 & 4400 & 4314 & 4427 & \\
& 2 & 4314 & 4434 & 4305 & 4440 & 4271 \\
$ns\bar{b}\bar{b}$ & 0 & 10721/10852 & 10999/11095 & 10955/11160 &
10883/11098 & 10802 \\
& \multirow{2}{*}{1} & 10684/10736 & 10911/11010 & 10811/10974 &
10734/10897 & \multirow{2}{*}{10706/10809} \\
& & 10778 & 11037 & 11068 & 11046 & \\
& 2 & 10747 & 11060 & 10997 & 10915 & 10823 \\
$ss\bar{c}\bar{c}$ & 0 & 4205/4474 & 4395/4672 & 4300/4521 &
4417/4587 & 4359 \\
& 1 & 4323 & 4526 & 4382 & 4493 & 4375 \\
& 2 & 4381 & 4597 & 4433 & 4536 & 4402 \\
$ss\bar{b}\bar{b}$ & 0 & 10868/10972 & 11157/11254 & 11078/11232 &
10972/11155 & 10932 \\
& 1 & 10908 & 11199 & 11099 & 10986 & 10939 \\
& 2 & 10926 & 11224 & 11119 & 11004 & 10950 \\
$ss\bar{c}\bar{b}$ & 0 & 7653/7833 & 7774/7960 & 7693/7875 &
7735/7894 & 7673 \\
& \multirow{2}{*}{1} & 7711/7764 & 7793/7872 & 7716/7757 &
7752/7775 & \multirow{2}{*}{7683/7684} \\
& & 7816 & 7924 & 7858 & 7881 & \\
& 2 & 7862 & 7908 & 7779 & 7798 & 7701 \\ \hline\hline
\end{tabular}%
\end{table*}

\renewcommand{\tabcolsep}{0.26cm} \renewcommand{\arraystretch}{1.1}
\begin{table*}[!htb]
\centering
\caption{Comparison of calculated mass (in MeV) with calculations conducted in DMC. ``NB" represents no bound solution. We only took the results which are bound state solutions from Ref. \cite{Ma:2023int}.}
\label{tab:comparisonDMC}%
\begin{tabular}{ccccccc}
\hline\hline
\textrm{State} & $J^P$ & \textrm{This\ work} & AL1 \cite{Ma:2023int} & AP1 \cite{Ma:2023int}& SLM \cite{Ma:2023int}  \\ \hline
${nn\bar{c}\bar{b}}^{I=0}$ & $0^+$ & 7108/7201 & 7136  & 7164 &6986 \\
${nn\bar{c}\bar{c}}^{I=0}$ & $1^+$ & 3892/4130 & 3880  & 3916 &3759 \\
${nn\bar{b}\bar{b}}^{I=0}$ & $1^+$ & 10338/10624 & 10500  & 10510 &10249\\
${ns\bar{b}\bar{b}}^{I=1/2}$ & $1^+$ & 10736/10778/10684 & 10660  & 10667 &10653\\
${nn\bar{c}\bar{b}}^{I=0}$ & $1^+$ & 7084/7324/7204 & 7200  & 7224 &7012\\
${nn\bar{c}\bar{b}}^{I=0}$ & $2^+$ & 7360 & 77367  & 7400 &NB\\
\hline\hline
\end{tabular}%
\end{table*}

Finally, we compare our results of magnetic moments with the available results in 
Table \ref{magneticmoment}. Our results are in general compatible with the cited work.

\renewcommand{\tabcolsep}{0.33cm} \renewcommand{\arraystretch}{1.1}
\begin{table*}[!htb]
\centering
\caption{Comparison of magnetic moments (in units of nuclear magneton $\protect\mu_{N}$, $\mu_p=2.79285\mu_N$).}
\label{magneticmoment}%
\begin{tabular}{cccccc}
\hline\hline
\textrm{State} & $J^{P}$ &   & Present work & $%
\mu$ \cite{Zhang:2021yul} \\ \hline
\multirow{2}{*}{${(nn\bar{c}\bar{c})}^{I=1}$} & \multirow{2}{*}{$0^{+}$} &
&  - & -  \\
&  &  &  -  & -  \\
& $1^{+}$ &  &   1.14, -0.07, 1.26  & 1.36, -0.03, -1.43 \\
& $2^{+}$ &  & 2.66, -0.06, 2.39   & 2.80, -0.05, -2.90 \\
\hline
\multirow{2}{*}{${(nn\bar{c}\bar{c})}^{I=0}$} & \multirow{2}{*}{$1^{+}$} &
 &   -0.28&  -0.88 &  \\
&  &  &  0.35 &  0.83 &  \\
\hline
\multirow{2}{*}{${(nn\bar{b}\bar{b})}^{I=1}$} & \multirow{2}{*}{$0^{+}$} &
&   - & -  \\
&  &  &  - & -  \\
& $1^{+}$ &  &   -1.02, 0.28, -0.64 & 1.81, 0.52, -0.78  \\
& $2^{+}$ &  &  -2.32, 0.47, -1.28& 3.65, 1.04, -1.57 \\ 
\hline
\multirow{2}{*}{${(nn\bar{b}\bar{b})}^{I=0}$} & \multirow{2}{*}{$1^{+}$} &
 &   0.32 & 0.18  & \\
&  &  &   0.69 & 0.86 &  \\
\hline
\multirow{2}{*}{${(nn\bar{c}\bar{b})}^{I=1}$} & \multirow{2}{*}{$0^{+}$} &
 &  - & -  \\
&  &  &  - & -  \\
& \multirow{3}{*}{$1^{+}$} &  &   1.84, 0.12, -1.36 & 2.33, 0.21, -1.90  \\
&  &  &  0.09, 0.12, 0.21& -0.17, -0.32, -0.47  &  \\
&  &  & 1.94, 0.41, -0.59 & 2.57, 0.82, -0.93 \\
& $2^{+}$ &  &   2.47, 0.32, -1.68  & 3.24, 0.50, -2.24  \\ 
\hline
\multirow{2}{*}{${(nn\bar{c}\bar{b})}^{I=0}$} & \multirow{2}{*}{$0^{+}$} &
 &   - & -  \\
&  & &   - & - \\
& \multirow{3}{*}{$1^{+}$} &  & 0.74  & 0.55
\\
&  &  &   0.70 &   0.50 \\
&  &  &   0.45 & -0.33   \\
& $2^{+}$ &  &   -0.54 & 0.50 \\ 
 \hline 
\multirow{2}{*}{$ns\bar{c}\bar{c}$} & \multirow{2}{*}{$0^{+}$} & & - & - \\
 &  & & - & - \\
& \multirow{3}{*}{$1^{+}$} & & 0.43, -0.67& 0.32, -1.33 \\
&   & & 1.24, -1.96 & 0.86, -1.58  \\
&   & & 0.52, -0.75 & -0.95,-1.03   \\
& $2^{+}$ && 0.24, -2.76  & 0.19, -2.68  \\\hline
\multirow{2}{*}{$ns\bar{b}\bar{b}$} & \multirow{2}{*}{$0^{+}$}  & &- & -  \\
&  & &- & - \\
& \multirow{3}{*}{$1^{+}$} &  & 0.48,
-0.66 & 0.65, -0.68  \\
&   & & 1.11, -1.35 & 1.02, -1.50 \\
&  & & -0.09, -0.12 & 0.14, 0.16   \\
& $2^{+}$ & & 1.36, -1.98 & 1.25, -1.39  \\\hline

\multirow{2}{*}{$ss\bar{c}\bar{c}$} & \multirow{2}{*}{$0^{+}$} &  & - & -  \\
 &  &&  -& - \\
& $1^{+}$ &   &-1.33 & -1.22  \\
& $2^{+}$ &  & -1.76 & -2.46 \\\hline
\multirow{2}{*}{$ss\bar{b}\bar{b}$} & \multirow{2}{*}{$0^{+}$} &  & - & -  \\
&  & & -& -  \\
& $1^{+}$ & & 0.73 & -0.60   \\
& $2^{+}$ & & 1.42 & -1.20  \\ \hline
\multirow{2}{*}{$ss\bar{c}\bar{b}$} & \multirow{2}{*}{$0^{+}$} &   & - & -  \\
&  & & - & -  \\
& \multirow{3}{*}{$1^{+}$} & & -0.86 & -1.54
\\
&  &   & -0.35 & -0.48  \\
&  &  & -0.21 & -0.66  \\
& $2^{+}$ &  & -2.09& -1.84  \\

\hline\hline
\end{tabular}%
\end{table*}

\section{\label{sec:level4}Concluding Remarks}
We employ a diffusion Monte Carlo (DMC) method to investigate systematically masses and magnetic moments of doubly heavy tetraquark states in constituent quark model. This method has some advantages compared to other methods; (i) reduces the uncertainty of the numerical procedure, (ii) accounts for multiparticle correlations, and (iii)  avoids quark clustering in the relevant states. 

Most QCD states arise as bound states or as resonances of multihadron configurations. Studying mass spectrum is an important aspect for the general view point of the related states. To obtain possible mass locations around the corresponding thresholds is not only crucial for understanding the underlying dynamics of the exotic hadrons and the nature of strong interactions of QCD but also useful for experimental researches for their
existence.

In the $nn \bar c \bar c$ systems,  the $I(J^P)=0(1^+)$ $nn \bar c \bar c$ state with mass 3892 MeV is the candidate of the experimentally observed $T_{cc}^{+}$ state. Its mass is only 16 MeV above $D D^\ast$ threshold.  $T_{cc}^{+}$ is argued to be a shallow bound state. Although the level of the uncertainty is less than 1 MeV, one cannot conclude this state is bound or not. Our results suggest $I(J^P)=0(1^+)$ $nn \bar c \bar c$ could be a shallow bound state. All the other predicted states with $I(J^P)=0(1^+)$ and $I(J^P)=1(1^+)$ for  $nn \bar c \bar c$ tetraquarks are well above corresponding thresholds. The magnetic moment presents information about the inner quark structure and can be used to distinguish the preferred configuration from predictions of different theoretical models. Our prediction of the magnetic moment for $T_{cc}^+$ state is $\mu=0.28 \mu_N$.

In the $nn \bar b \bar b$ systems, we find a deeply bound state candidate, namely $I(J^P)=0(1^+)$ $nn \bar b \bar b$ state with mass 10338 MeV. It is just 266 MeV below $\bar{B} \bar{B}^\ast$ threshold. All the other $nn \bar c \bar c$ states are above their corresponding thresholds and could be resonances. The existence  $T_{bb}^-$ was conjectured long time ago and recent lattice QCD studies confirm the existence of $ud\bar b \bar b$ tetraquark with $I(J^P)=0(1^+)$ \cite{Francis:2016hui,Junnarkar:2018twb,Leskovec:2019ioa,Mohanta:2020eed}. We obtain magnetic moment of $T_{bb}^-$ as $\mu=-0.32 \mu_N$. 

In the $nn \bar c \bar b$ systems, we observe some bound state candidates. For $I=0$, low-lying $J^P=0^+$ $nn \bar c \bar b$ has a mass of $7108$ MeV and is 39 MeV below $D \bar{B}$ threshold. One of the $J^P=1^+$ $nn \bar c \bar b$ has a mass of 7084 MeV and is 108 below $D \bar{B}^\ast$ threshold and 204 MeV below $D^\ast \bar{B}^\ast$ threshold. The $J^P=1^+$ $nn \bar c \bar b$ state with mass of 7204 MeV is 12 MeV above $D \bar{B}^\ast$ threshold and 84 MeV below $D^\ast \bar{B}^\ast$ threshold. The two low-lying 
$nn \bar c \bar b$ states with $J^P=1^+$ could be a bound or loosely bound state with regard to corresponding threshold. Some of the recent lattice QCD calculations \cite{Hudspith:2020tdf,Pflaumer:2021ong,Meinel:2022lzo} did not find any bound state indications of $I(J^P)=0(0^+)$ and $I(J^P)=0(1^+)$ $ud \bar{c} \bar{b}$ states but cannot rule out their existence. However, Refs. \cite{Francis:2018jyb,Mathur:2021gqn} support $I(J^P)=0(1^+)$ $ud \bar{c} \bar{b}$ existence.  A very recent lattice QCD study also found evidence for $J=0$ shallow bound state and found hints for $J=1$ broad resonance of $ud \bar{c} \bar{b}$ \cite{Alexandrou:2023cqg}.

In the $ns \bar c \bar c$ and $ns \bar b \bar b$ systems, we find all the $ns \bar c \bar c$ states above corresponding thresholds. They could be resonances. In the case of $ns \bar b \bar b$ systems, the low-lying $J^P=1^+$ state is 8 MeV below $\bar{B}^\ast \bar{B}_s$ threshold and 11 MeV below $\bar{B}_s^\ast \bar{B}$ threshold. This could be a bound state candidate. The 
$J^P=2^+$ state is 7 MeV above $\bar{B}^\ast \bar{B}_s^\ast$ threshold. This  could be a shallow bound state candidate. In Ref. \cite{Meinel:2022lzo}, a clear evidence for the existence of $us\bar b \bar b$ with $J^P=1^+$ is found. 

Lattice QCD studies present important evidences about the existence of the states. For the $I=0$ $T_{bb}$, $I=0,1$ $T_{bc}$ and $T_{bbs}$ tetraquark states, lattice QCD results support deeply bound state or very narrow resonances (see a recent review on lattice QCD results and references therein  \cite{Bicudo:2022cqi}. Our results are in agreement with those predictions.

In the $ss \bar c \bar c$, $ss \bar b \bar b$, and $ss \bar c \bar b$ systems, all the states are above their corresponding thresholds. They all could be resonances. We have no bound state candidates in these systems.

The experimental progress on the multiquark states open a new era for our understanding of QCD. We hope that our results will be useful for not only for experimental studies but also other theoretical studies.

\bibliography{QQqqDMC2}

\end{document}